\documentclass[12pt,english,sort&compress]{article}
\usepackage[latin9]{inputenc}
\usepackage{amsmath}
\usepackage{graphicx}
\usepackage[numbers]{natbib}
\usepackage{microtype}
\usepackage{textcomp}

\makeatletter

\newcommand{\lyxaddress}[1]{
	\par {\raggedright #1
	\vspace{1.4em}
	\noindent\par}
}

\usepackage{aecompl}
\usepackage{amsmath}
\usepackage{textcomp}
\usepackage{float}

\usepackage{aecompl}\usepackage{epstopdf}
\usepackage{babel}

\makeatother

\usepackage{babel}
\begin{document}
\date{\empty}
\title{Line and planar defects with zero formation free energy: Applications of the phase rule toward ripening-immune microstructures}
\author{Ju Li$^{1,2}$ and Yuri Mishin$^{3}$}
\maketitle

\lyxaddress{$^{1}$ Department of Nuclear Science and Engineering, Massachusetts
Institute of Technology, Cambridge, MA 02139, USA}

\lyxaddress{$^{2}$ Department of Materials Science and Engineering, Massachusetts
Institute of Technology, Cambridge, MA 02139, USA}

\lyxaddress{$^{3}$ Department of Physics and Astronomy, MSN 3F3, George Mason
University, Fairfax, Virginia 22030, USA}

\begin{abstract}
Extended one- and two-dimensional defects in crystalline
materials are usually metastable. The thermodynamic
ground state of the material is presumed to be defect-free. Here,
we investigate the conditions under which extended defects, such as grain boundaries, can exist
in a multicomponent alloy when the latter reaches the thermodynamic ground state allowed by the Gibbs phase rule. We treat all extended
defects as low-dimensional phases on the same footing as the conventional
bulk phases. Thermodynamic analysis shows that, in the ground state,
the formation free energies of all extended defects must be zero, and
the system must follow a generalized phase rule. The latter predicts
that only a finite number of symmetry-related defect types can coexist
in the material in the ground state. Guided by the phase rule,
we discuss finite-size polycrystalline and/or polyphase microstructures that are fully immune to coarsening and their possible transformations.
\end{abstract}

\section{Introduction}

Most crystalline solids contain defects, which are traditionally classified
into the following categories: 
\begin{itemize}
\item Zero-dimensional (0D) defects, also known as point defects, include
vacancies, self-interstitials, impurity atoms, and similar disruptions
of crystalline order with atomic-scale size in all directions. 
\item One-dimensional (1D) defects, including dislocations, triple junctions
of grains in polycrystalline materials, contact lines between phases,
and other defects with atomic-scale dimensions in two directions and mesoscale length along a line. 
\item Two-dimensional (2D) defects, such as grain boundaries, interphase
boundaries, stacking faults, and other defects of mesoscale area and
atomic-scale width. 
\end{itemize}
3D defects can also be considered, such as pores and stacking-fault
tetrahedra, but they are not discussed in this paper. The 1D and 2D
defects are collectively referred to as extended defects.

Intrinsic point defects, such as vacancies and self-interstitials,
are classified as \emph{equilibrium defects} because they are part
of the thermodynamic ground-state structure of crystalline phases.
At a given temperature $T$, they reach equilibrium concentrations
\begin{equation}
c^{{\rm eq}}(T)\propto\exp(-f_{{\rm 0D}}/k_{{\rm B}}T),\label{eq:s-s}
\end{equation}
where $f_{{\rm 0D}}$ is the local (nonconfigurational) formation
free energy per defect and $k_{{\rm B}}$ is Boltzmann's constant.
The intrinsic point defects form spontaneously to reduce the Helmholtz
free energy of the solid by increasing its configurational entropy.
The latter is associated with different positions of the defects on
the crystalline lattice, which they sample by a diffusive random walk.

In contrast, most extended defects are thermodynamically unstable
or metastable. They have an excess nonconfigurational free energy
and seldom contribute to the configurational entropy because their
mobility is too low, preventing them from sampling the positions
in space on the timescale of laboratory experiments and practical
applications. As a result, their formation free energy is dominated by the
non-configurational part. Its value per unit length
$L$ or per unit area $A$ is often referred to as, respectively,
the free energy of the line $\tau$ and the free energy of the surface/interface $\gamma$.
These quantities are sometimes called line tension and surface tension, respectively,
if directional isotropy
is assumed (e.g., fluid/fluid interfaces). An extended defect cannot form spontaneously by thermal
fluctuations because its total formation free energy is exorbitantly
high. For example, the total free energy $F=\tau L$ of a dislocation
line of length $L$ (typically measured in micrometers) can be on
the order of keV or higher. As a result, the thermally equilibrium
population of dislocations, estimated from the Boltzmann factor $\propto\exp{(-F/k_{{\rm B}}T)}$,
is negligibly small. The same applies to the thermally equilibrium
population of grain boundaries.

Extended defects can lower their length/area spontaneously if they
possess sufficient mobility. They can also increase their length/area
if an external driving force is applied, such as shear stress, radiation,
or temperature change. For example, a Peach-Koehler force acting on
a dislocation can cause it to bow out and increase its length. In
many cases, extended defects are found in a stationary state when
they are pinned by immobile elements of the microstructure. Stationary
defects can reach a state of \emph{constrained} thermodynamic equilibrium,
i.e., equilibrium with respect to exchanges of heat and chemical components
with the environment. In alloy systems, such exchanges lead to the
formation of solute segregation at interfaces and segregation atmospheres
(Cottrell atmospheres) at dislocations. The solute segregation reduces
the free energy of defect formation according to the Gibbs adsorption equation for multicomponent systems \cite{Willard_Gibbs}, but the latter remains positive
in most cases. We emphasize that the defect is \emph{not} in full
thermodynamic equilibrium because its length/area is not allowed to
vary to further reduce the free energy.

It was suggested \citep{Weissmuller:1992aa,Weissmuller:1993aa,Weissmuller1994,Krill-1995,Kirchheim:2002aa,Liu:2004aa,Krill:2005aa,Schvindlerman2006,Kirchheim2007a,Kirchheim2007b,Detor2007,Trelewicz2009,Chookajorn2012,Murdoch:2013aa,Chookajorn2014,Kalidindi:2017aa,Kalidindi:2017bb,Kalidindi:2017cc,Perrin-2021,Hussein:2024aa,Hussein:2025aa}
that a sufficiently strong grain boundary segregation of solute atoms
can fully stabilize a polycrystalline alloy against grain growth.
Thermodynamic analysis shows that the total free energy of a closed
system composed of a uniform grain boundary and surrounding grains
can reach a minimum with respect to variations of the boundary area
when the boundary free energy $\gamma$ reaches zero value \citep{Hussein:2024aa}.
The derivation assumes that during the area variations, the grain
boundary remains in thermal and chemical equilibrium with the grains.
Extensive experimental, theoretical, and computational efforts \citep{Detor2007,Chookajorn2012,Chookajorn2014,Kalidindi:2017aa,Kalidindi:2017bb,Kalidindi:2017cc,Murdoch:2013aa,Trelewicz2009,Perrin-2021,Koch2008,Darling2014,Saber2013,Saber2013a,Xing:2018aa,Zhou:2014aa,Hussein:2024aa,Hussein:2025aa}
have been dedicated to the search for solutes that could drastically
reduce $\gamma$ and even drive it to a zero value. However, the thermodynamic meaning of the conditions of $\gamma=0$ or $\tau=0$ and their microstructural consequences have not been systematically discussed.

Here, we suggest that the thermodynamics underlying the grain boundary
stabilization is as general as the thermodynamics of 3D phases, in which multiple 3D phases can coexist in equilibrium if certain thermodynamic conditions are met. Thermodynamics is equally valid for all 2D and
all 1D defect phases in crystalline materials. For 1D defects, the $\gamma=0$
condition must be replaced with $\tau=0$. A truly stabilized material
must satisfy the $\gamma=0$ and $\tau=0$ conditions for \emph{all} 2D and
1D defects present in the material. For example, if all grain boundaries
in a polycrystalline material are fully stabilized ($\gamma=0$) but
$\tau$ of triple junctions remains positive, the defect structure
will still remain unstable and evolve under the capillary forces of
the triple junctions. The total free energy will continue to strive
towards smaller values.

The full thermodynamic stabilization of crystalline materials is an
important fundamental concept, but it also presents significant practical
interest. The pursuit of full thermodynamic stabilization of polycrystals
and/or finite-size phases (e.g., phase precipitates) that are forever immune to Ostwald ripening or coarsening opens a
previously unrecognized design space, particularly for
nanocrystalline materials whose superior physical and mechanical properties
predicate on the suppression of grain growth at elevated temperatures.
The full thermodynamic stabilization, if achieved, will suppress all
driving forces while retaining the nanostructure responsible
for the superior properties. The full stabilization concept is very
general and relies only on the laws of thermodynamics. If realized
in practice, it would circumvent the currently employed kinetic stabilization
mechanisms such as solute drag and Zener pinning by small embedded
particles.

The central idea of this article is that extended defects can be treated
as 1D and 2D phases on par with conventional bulk (3D) phases. Thus,
the material can be considered to be composed of multiple phases of
different dimensionality: 1D and 2D phases of extended defects and
3D bulk phases. Generally, these phases are not in equilibrium with
each other. However, the system can reach the thermodynamic ground
state if all phases establish equilibrium with each other. It can
be shown (see Supplementary Information) that the condition
of $\gamma=0$ and $\tau=0$ for all defects is equivalent to achieving
the thermodynamic equilibrium between all phases present in the system.
This insight is crucial because it allows us to apply the Gibbs phase
rule to the fully stabilized crystalline material containing defects.
Depending on the number of chemical components, the phase rule can
predict the maximum numbers of different 1D and 2D defects that can
coexist in the system when it reaches the ground state. These numbers
inform the analysis of possible equilibrium microstructures of a fully
stabilized material, depending on the dimensionalities of the defects
and the crystal symmetry. Such microstructures could be designed to
achieve the desired physical and mechanical properties while simultaneously
preventing them from coarsening at high temperatures.

\section{Thermodynamics of extended defects and the phase rule}

We consider a $k$-component alloy composed of several bulk phases
and grains. The alloy contains interfaces and line defects. The extended
defects are treated as phases on equal footing with the bulk phases.
For brevity, we will use the shorthand $(\varphi_{3}@\varphi_{2}@\varphi_{1})_{k}$
for a $k$-component system composed of $\varphi_{3}$ bulk phases,
$\varphi_{2}$ 2D phases, and $\varphi_{1}$ 1D phases. For example,
$(1@1@0)_{2}$ denotes a single-phase binary solid solution containing
a grain boundary.

Thermodynamic properties of a bulk phase are fully defined by a fundamental
equation 
\begin{equation}
F=F(T,N_{1},...,N_{k},V),\label{eq:50}
\end{equation}
where $F$ is the total free energy of the phase, $T$ is temperature,
$V$ is volume, and $N_{i}$ are the amounts of the chemical components
(in moles). The phase is considered spatially uniform and thermally
equilibrated (uniform temperature throughout). Under these assumptions,
the free energy is a homogeneous first-degree function of $V$ and
all $N_{i}$'s. Applying Euler's theorem of homogeneous functions,
we obtain 
\begin{equation}
F=-pV+\sum_{i=1}^{k}\mu_{i}N_{i},\label{eq:51}
\end{equation}
where $\mu_{i}=\partial F/\partial N_{i}$ are the chemical potentials
of the components and $p=-\partial F/\partial V$ is pressure.

Next, consider an interface between two phases or two grains in the
same phase. All thermodynamic properties of the interface can be derived
from its fundamental equation \citep{Frolov:2015ab,Mishin:2015ab}
\begin{equation}
\tilde{F}=\tilde{F}(T,\tilde{N}_{1},...,\tilde{N}_{k},A),\label{eq:52}
\end{equation}
where $A$ is the interface area, $\tilde{F}$ is the excess free
energy of the interface, and $\tilde{N}_{i}$ are the excess amounts
of chemical components (in moles). In Eq.(\ref{eq:52}), the excesses
are defined using Gibbs' dividing surface construction \citep{Gibbs}.
Namely, the excess quantity is calculated relative to homogeneous
bulk phases by extrapolating their intensive properties to the geometric
dividing surface.  There are many ways to define
interface excess properties \citep{Frolov:2015ab,Cahn79a}, which
all lead to the same final results. In a more general treatment \cite{Frolov:2015ab}, the interface free energy is expressed in terms
of generalized excesses introduced by Cahn \cite{Cahn79a}. For this discussion, it will suffice to adopt one particular type of excess, namely excess under the constraint of fixed volume. Such excesses are denoted $[X]_V$ ($X$ being any extensive property) \cite{Frolov:2015ab} and correspond to the dividing surface construction introduced by Gibbs \citep{Gibbs}. Note that, under this convention, the excess volume of any interface is zero by definition. All equations appearing in the following can be reformulated in terms of other definitions of excesses without changing the conclusions.

Note that Eq.(\ref{eq:52}) has the same functional
form as Eq.(\ref{eq:50}), except that the spatial dimension of the
interface is defined by its area $A$ instead of the volume $V$.
The interface properties are assumed to be uniform throughout the
area. Therefore, the excess free energy in Eq.(\ref{eq:52}) is again
a homogeneous function of first degree with respect to the extensive
variables $A$ and $\tilde{N}_{i}$'s. From Euler's theorem, 
\begin{equation}
\tilde{F}=\gamma A+\sum_{i=1}^{k}\tilde{\mu}_{i}\tilde{N}_{i}.\label{eq:53}
\end{equation}
Here, 
\begin{equation}
\gamma=\dfrac{\partial\tilde{F}}{\partial A}\label{eq:54}
\end{equation}
is the interface free energy (tension) and we have introduced the notation
$\tilde{\mu}_{i}\equiv\partial\tilde{F}/\partial\tilde{N}_{i}$. Note
that $\gamma$ is a 2D analog of the (negative) pressure $p$.

Similarly, all thermodynamic properties of a line defect can be derived
from the fundamental equation \citep{Frolov:2015ab,Mishin:2015ab}
\begin{equation}
\hat{F}=\hat{F}(T,\hat{N}_{1},...,\hat{N}_{k},L),\label{eq:55}
\end{equation}
where $L$ is the length of the defect and the excess quantities are calculated
by extrapolating the intensive properties of the homogeneous bulk
phases to the defect line. The excess free energy in Eq.(\ref{eq:55})
is a homogeneous function of first degree similar to Eq.(\ref{eq:50})
except for the replacement of $V$ by $L$. As above, we apply Euler's
theorem to obtain 
\begin{equation}
\hat{F}=\tau A+\sum_{i=1}^{k}\hat{\mu}_{i}\hat{N}_{i},\label{eq:56}
\end{equation}
where $\hat{\mu}_{i}\equiv\partial\hat{F}/\partial\hat{N}_{i}$ and
\begin{equation}
\tau=\dfrac{\partial\hat{F}}{\partial L}\label{eq:57}
\end{equation}
is the defect free energy (line tension), which is a 1D analog of
the interface tension $\gamma$.

The similarity of the fundamental equations (\ref{eq:50}), (\ref{eq:52})
and (\ref{eq:55}) justifies treating the extended defects as phases.
As indicated earlier \citep{Frolov:2015ab}, the laws of thermodynamics
are not specific to any particular dimension of space. Any object
whose thermodynamic properties are described by a fundamental equation
can be treated as a phase. Thermodynamically, a 3D phase is no different
from 1D/2D defect phases or other low-dimensional systems such as
suspended graphene, graphane, or twisted MoS$_{2}$ bilayer.

The total free energy of a multiphase, multidefect system is the
sum of the free energies of all phases present in the system in the
form of equations (\ref{eq:50}), (\ref{eq:52}) and (\ref{eq:55}).
Suppose that the phases are initially not in equilibrium with each other.
Then we bring them to thermal, mechanical, and chemical equilibrium
with each other. As discussed in the Supplementary Information file,
the equilibrium conditions can be readily derived by considering reversible
variations of the entropies of all phases and volumes of the bulk
phases, as well as exchanges of the chemical components between the
phases. The result is that the temperatures of all phases must be
equal, the pressures in the bulk phases must be equal, and the coefficients
in front of $\tilde{N}_{i}$ and $\hat{N}_{i}$ in Eqs.(\ref{eq:53})
and (\ref{eq:56}) must be equal to the respective chemical potentials:
$\hat{\mu}_{i}=\tilde{\mu}_{i}=\mu_{i}$ for all $i=1,...,k$. In
other words, the chemical potential of each species is the same in
all phases of any dimensionality. Under these conditions, Eqs.(\ref{eq:53})
and (\ref{eq:56}) become 
\begin{equation}
\tilde{F}=\gamma A+\sum_{i=1}^{k}\mu_{i}\tilde{N}_{i},\label{eq:58}
\end{equation}
\begin{equation}
\hat{F}=\tau L+\sum_{i=1}^{k}\mu_{i}\hat{N}_{i}.\label{eq:59}
\end{equation}

The above conditions of thermal, mechanical, and chemical equilibrium
are not sufficient for reaching the thermodynamic ground state. They
only bring the system to a state of constrained equilibrium, in which
the segregation atmospheres on extended defects are in equilibrium
with the bulk phases, but the defects are not allowed to alter their
areas and lengths. To reach the ground state, the free energy must
be minimized with respect to variations of the defects' areas and
lengths. The result of this minimization (see Supplementary Information
file) is that in the ground state, the formation free energies of
all extended defects must be zero: 
\begin{equation}
\tau_{1}=\tau_{2}=...=\tau_{\varphi_{1}}=0,\label{eq:60}
\end{equation}
\begin{equation}
\gamma_{1}=\gamma_{2}=...=\gamma_{\varphi_{2}}=0.\label{eq:61}
\end{equation}
The possible defect configurations in the ground state will be discussed
later.

The system can remain in the ground state while some of its thermodynamic
parameters vary without changing the number of phases. The number
$\pi$ of independent parameters that can vary (the number of degrees
of freedom) can be derived by subtracting the total number of constraints
imposed by the equilibrium conditions from the total number of intensive
parameters describing the system. The calculation gives the following
phase rule for a crystalline material with defects (see Supplementary
Information file): 
\begin{equation}
\pi=k+2-(\varphi_{1}+\varphi_{2}+\varphi_{3}).\label{eq:62}
\end{equation}

For example, consider a $(1@1@0)_{2}$ system composed of a binary
solid solution with a grain boundary (or a set of symmetrically equivalent
grain boundaries) at $\gamma=0$. Eq.(\ref{eq:62}) predicts $\pi=2$.
If the pressure is fixed, the system has one degree of freedom. For
example, we can vary the temperature, which will cause changes in
grain boundary segregation, the chemical composition of the grains,
and the grain boundary area. If we also fix the temperature, then
there will be no degrees of freedom left. We can still vary the average
chemical composition of the alloy by adding or removing the solute
atoms. This will not cause any changes in the chemical composition
inside the grains or at the grain boundary. However, the grain boundary
area will vary to accommodate the added/removed solute atoms. The
same solid solution can be equilibrated with two different grain boundaries
(assuming that they represent different 2D phases), but only at one
temperature.

As another example, consider a tri-crystal (three grain boundaries
and a triple line) in a five-component solid solution, which is a
$(1@3@1)_{5}$ system. When this system reaches a thermodynamic ground
state, it has $\pi=2$ degrees of freedom. At a fixed pressure, the
temperature can be varied while keeping the system in the ground state.
Temperature variations will be accompanied by changes in grain composition
and boundary/junction segregations with corresponding adjustments
of their areas/lengths.

\begin{figure}[t]
\begin{centering}
\includegraphics[width=0.42\textwidth]{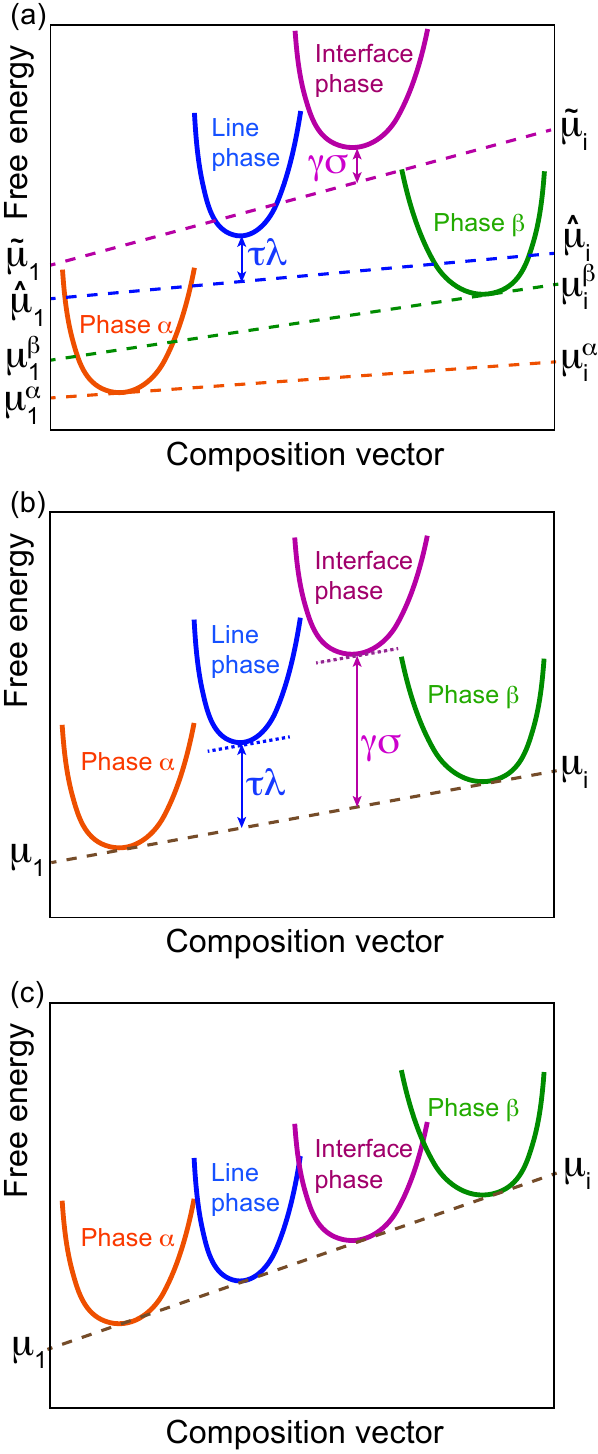} 
\par\end{centering}
\caption{Conceptual diagrams illustrating the geometry of the defect phase
thermodynamics. The molar free energy is plotted against the $(k-1)$-dimensional
composition vector. The multicomponent $(2@1@1)_{k}$ alloy contains
two bulk (3D) phases $\alpha$ and $\beta$, a 1D defect phase (line
defect), and a 2D defect phase (interface). (a) The phases are not
in equilibrium with each other. (b) The system is in constrained equilibrium
(the chemical potentials in all phases are equal). (c) The system
is in thermodynamic ground state. The dashed lines represent the $\Lambda$
-planes. The dotted lines in (b) represent tangential planes to the
free energies of defect phases.}
\label{fig:diagrams-A} 
\end{figure}

We next discuss a geometric interpretation of the defect thermodynamics.
It is convenient to reformulate the fundamental equations in intensive
variables. For a bulk phase, we introduce the molar free energy $f=F/N$
and the concentrations (mole fractions) of the components $c_{i}=N_{i}/N$,
where $N=\sum_{i}N_{i}$ is the total amount of the chemical components.
The chemical composition is specified by $(k-1)$ independent concentrations,
for which we choose $c_{2},...,c_{k}$. These concentrations form
a $(k-1)$-dimensional composition vector $\mathbf{C}=(c_{2},...,c_{k})$.
Eqs.(\ref{eq:50}) and (\ref{eq:51}) become 
\begin{equation}
f=f(T,c_{2},...,c_{k},\omega),\label{eq:63}
\end{equation}
\begin{equation}
f=-p\omega+\left(\mu_{1}+\sum_{i=2}^{k}\mu_{i1}c_{i}\right),\label{eq:64}
\end{equation}
where $\omega=V/N$ is the molar volume and $\mu_{i1}=\mu_{i}-\mu_{1}$
are the diffusion potentials relative to reference component 1. Similarly,
for the excess free energy of an extended defect we have

\begin{equation}
\tilde{f}=\tilde{f}(T,\tilde{c}_{2},...,\tilde{c}_{k},\sigma),\label{eq:65}
\end{equation}
\begin{equation}
\hat{f}=\hat{f}(T,\hat{c}_{2},...,\hat{c}_{k},\lambda),\label{eq:66}
\end{equation}
\begin{equation}
\tilde{f}=\gamma\sigma+\left(\tilde{\mu}_{1}+\sum_{i=2}^{k}\tilde{\mu}_{i1}\tilde{c}_{i}\right),\label{eq:67}
\end{equation}
\begin{equation}
\hat{f}=\tau\lambda+\left(\hat{\mu}_{1}+\sum_{i=2}^{k}\hat{\mu}_{i1}\hat{c}_{i}\right).\label{eq:68}
\end{equation}
Here, the excess free energy and the segregated amounts of components
have been normalized by the total excess amounts $\tilde{N}=\sum_{i}\tilde{N}_{i}$
and $\hat{N}=\sum_{i}\hat{N_{i}}$, respectively. Other notations
include the normalized area $\sigma=A/\tilde{N}$, the normalized
length $\lambda=L/\hat{N}$, and the low-dimensional analogs of the
diffusion potentials: 
\begin{equation}
\tilde{\mu}_{i1}=\dfrac{\partial\tilde{f}}{\partial\tilde{c}_{i}}-\dfrac{\partial\tilde{f}}{\partial\tilde{c}_{1}},\label{eq:69}
\end{equation}
\begin{equation}
\hat{\mu}_{i1}=\dfrac{\partial\hat{f}}{\partial\hat{c}_{i}}-\dfrac{\partial\hat{f}}{\partial\hat{c}_{1}}.\label{eq:70}
\end{equation}
The excess concentrations $\tilde{c}_{i}$ and $\hat{c}_{i}$ can
be grouped in composition vectors $\tilde{\mathbf{C}}$ and $\hat{\mathbf{C}}$,
respectively.

Consider a $k$-dimensional space spanned by the molar free energies
of the phases and the composition vectors (Fig.\ \ref{fig:diagrams-A}).
The terms in parentheses in Eqs.(\ref{eq:64}), (\ref{eq:67}) and
(\ref{eq:68}) define $(k-1)$-dimensional hyperplanes, which we call
$\Lambda$-planes because they originate from Legendre transformations
of the free energy with respect to the concentrations. The $\Lambda$-planes
are parallel to the tangential hyperplanes to the respective free
energy functions $f(\mathbf{C})$, $\tilde{f}(\tilde{\mathbf{C}})$
and $\hat{f}(\hat{\mathbf{C}})$ at fixed $T$, $\sigma$, and $\lambda$.
For a bulk phase, the $\Lambda$-plane coincides with the tangential
hyperplane (assuming $p=0$). For a defect phase, the $\Lambda$-plane
must be shifted by a positive amount of $\gamma\sigma$ or $\tau\lambda$
to touch the plot of the respective free energy function. In the initial
state, when the phases are not in equilibrium with each other, their
$\Lambda$-planes are generally not parallel. In the example shown
in Fig.\ \ref{fig:diagrams-A}(a), the bulk phases $\alpha$ and
$\beta$ are not in equilibrium with each other. If the free energy
$f_{\beta}(\mathbf{C})$ of phase $\beta$ is above the $\Lambda$-plane
of phase $\alpha$, then phase $\beta$ is metastable relative to
phase $\alpha$. In fact, phase $\beta$ can be considered a defect
in phase $\alpha$.

If the defect phases are in constrained equilibrium with each other
and with the bulk phases, then all $\Lambda$-planes merge into a
single hyperplane with linear coefficients equal to the diffusion
potentials (Fig.\ \ref{fig:diagrams-A}(b)). The bulk phases are
now in thermodynamic equilibrium with each other and their tangential
planes merge into a common tangent plane. However, if the formation
free energies $\gamma$ and $\tau$ of the defect phases remain positive,
the defects are metastable relative to the bulk phases. They form
equilibrium segregation atmospheres, but their existence costs the
system extra free energy. When the system finally reaches its ground
state, the formation free energies of the extended defects become
zero. All $\Lambda$ planes and all tangential planes merge into a
common tangent plane to all phases (Fig.\ \ref{fig:diagrams-A} (c)).

\begin{figure}[t]
\begin{centering}
\includegraphics[width=0.42\textwidth]{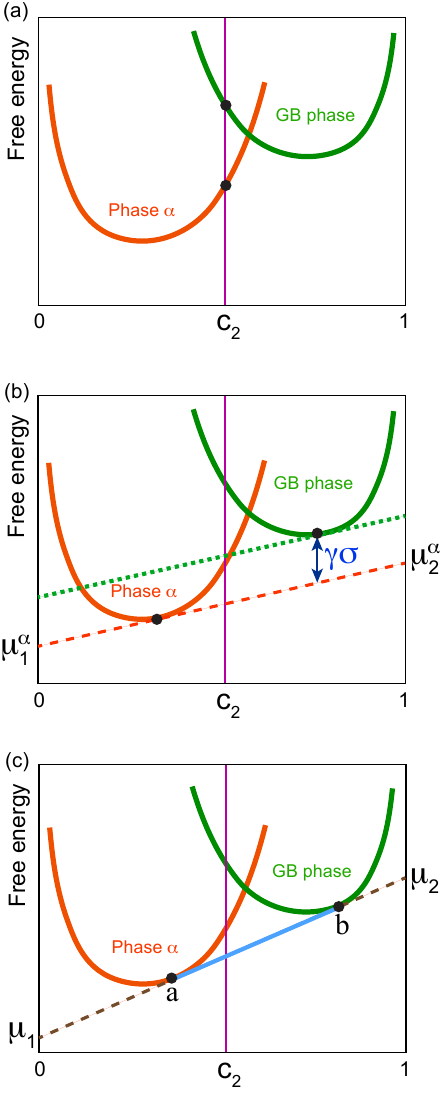} 
\par\end{centering}
\caption{Free energy diagram of a binary solid solution $\alpha$ containing
a grain boundary labeled ``GB phase''. The molar free energy is plotted
against the solute concentration $c_{2}$. (a) Initial state after
the boundary was inserted in the solution. (b) Grain boundary has
been equilibrated with the $\alpha$ phase without changing its area.
(c) The grain boundary area has been adjusted to achieve thermodynamic
equilibrium with the $\alpha$ phase. The vertical line marks the
fixed alloy composition. The dashed lines show the tangents to the
phases. The black dots indicate the current states of the phases.
The blue line $a-b$ is a tie line between the two phases.}
\label{fig:diagrams-B} 
\end{figure}

It is instructive to consider in more detail the case of a single
grain boundary in a binary solid solution, which is a $(1@1@0)_{2}$
system with $\pi=2$. Suppose the system is closed and the temperature
and pressure are fixed ($p=0$). The molar free energy of the bulk
phase, which we call phase $\alpha$, is a function of the solute
concentration $c_{2}$. The bulk solution is initially in internal
equilibrium. Next, we create a grain boundary in the system. Before
it has a chance to equilibrate with the environment, it has the same
chemical composition as the bulk phase (Fig.\ \ref{fig:diagrams-B}(a));
all excess properties are zero. Then we allow the boundary to equilibrate
with the $\alpha$ phase without changing the boundary area. This
can be achieved by diffusion of the solute atoms. The boundary forms
an equilibrium segregation atmosphere of the solute atoms. The process
is accompanied by redistribution of the solute between the grain boundary
and the bulk solution until a constrained equilibrium is reached.
If the segregation is positive ($\tilde{c_{2}}>0$), the bulk solution
is slightly depleted in the solute. The new chemical compositions
of the boundary and the solution are such that the tangents to the
respective free energy plots are parallel and the lever rule is satisfied
for the given grain boundary area (Fig.\ \ref{fig:diagrams-B}(b)).
The tangent line to the grain boundary free energy is a distance $\gamma\sigma$
above that of the $\alpha$ phase, indicating that the grain boundary
phase is metastable. Finally, we allow the grain boundary to decrease
its area to further reduce the total free energy. In the scenario
shown in Fig.\ \ref{fig:diagrams-B}(c), the system reaches a ground
state at a finite grain boundary area. The chemical compositions of
the phases and the finial grain boundary area are determined by the
common tangent construction and the lever rule applied to the tie
line $a-b$.

\begin{figure}[t]
\begin{centering}
\includegraphics[width=0.42\textwidth]{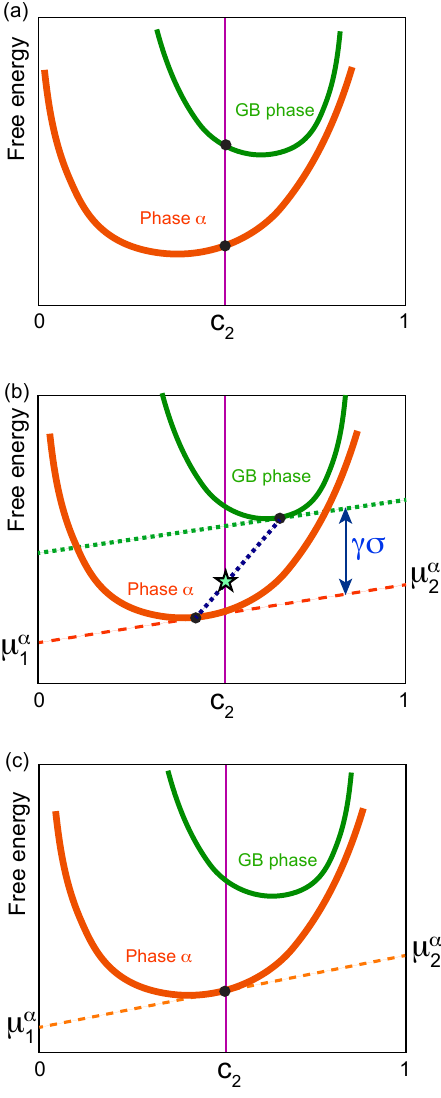} 
\par\end{centering}
\caption{Free energy diagram of a binary solid solution $\alpha$ containing
a grain boundary (GB phase). The molar free energy is plotted against
the solute concentration $c_{2}$. (a) Initial state after the boundary
was inserted in the solution. (b) Grain boundary has been equilibrated
with the $\alpha$ phase without changing its area. (c) The grain
boundary has disappeared to let the system reach the thermodynamic
ground state. The vertical line marks the fixed alloy composition.
The dashed lines show the tangents to the phases. The black dots indicate
the current states of the phases. The dotted line in (b) is a tie
line between the phases, and the star symbol marks the free energy
of the system.}
\label{fig:diagrams-C} 
\end{figure}

Fig.\ \ref{fig:diagrams-C} presents an alternative scenario. In
this case, the grain boundary again reaches a constrained equilibrium
with the environment at a fixed area (Fig.\ \ref{fig:diagrams-C}(b)).
As above, the new chemical compositions of the phases are dictated
by the parallel tangent construction and the lever rule. The star
symbol on the plot represents the total free energy of the system
in this metastable state. If the grain boundary is allowed to vary
its area, it will shrink and eventually disappear. The system will
return to its initial single-phase state, in which the free energy
is lower than in any state containing the grain boundary (Fig.\ \ref{fig:diagrams-C}(c)).
This example demonstrates that full stabilization of a grain boundary
is not always possible. It requires particular shapes of the free
energy functions of both the grain boundary and the grains. Note that
these shapes depend on temperature. A full grain boundary stabilization
can be possible at one temperature but may become impossible at another
temperature. In other words, a fully stabilized grain boundary can
exist in a certain domain of the composition-temperature phase diagram
of the bulk phase.

\section{Fully stabilized defect structures}

We next discuss possible morphologies of extended defects in a fully
stabilized state. We focus on grain boundaries in a single-phase multicomponent
solid solution as an example, although our conclusions are more general.

Grain boundary properties depend on five directional degrees of freedom
in addition to internal variables, such as the local atomic density
\citep{Frolov2013,Hickman__2017a}. Grain boundaries can also undergo
structural \citep{Frolov2013,Frolov:2018aa,Meiners:2020aa} and segregation-induced
\citep{Frolov:2015aa} phase transformations. Partitioning into families
of symmetry-related structures reduces the number of distinct grain
boundary types. However, we still have an infinite number of boundaries
that can potentially reach a full thermodynamic equilibrium. Meanwhile,
the phase rule dictates that only a finite (usually small) number
of grain boundary types can exist in the thermodynamic ground state.
How can nature reconcile the phase rule with the infinite pool of
grain boundary types? Furthermore, the full equilibration is achieved
for a specific grain boundary area and without triple junctions. What
kind of geometric arrangements of grain boundaries can satisfy these
requirements?

\begin{figure}[t]
\begin{centering}
\includegraphics[width=0.96\textwidth]{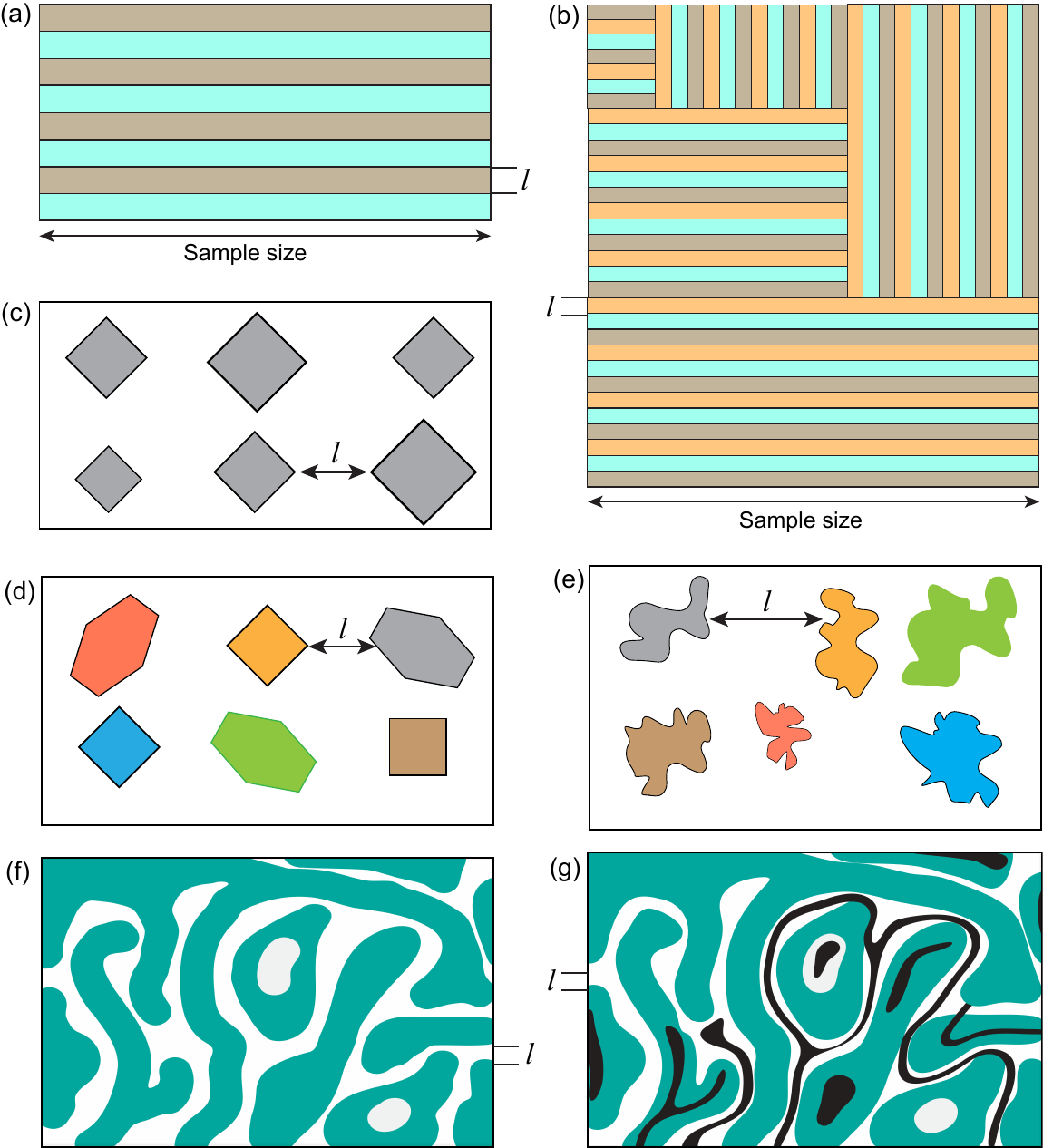} 
\par\end{centering}
\caption{2D schematics of possible ground state morphologies of a solid solution
with grain boundaries ($(1@1@0)_{k}$ system). (a) Lamellar structure
with symmetrical tilt grain boundaries. (b) Polysynthetic structure
composed of lamellar domains. (c) Isolated grain arrangement for a
single type of grain boundaries ($\xi=1$). (d) Isolated grain arrangement
for several grain boundary types ($\xi>1$). (e) Isolated grains above
the roughening transition. (f) Bi-continuous bicrystalline structure.
(g) Tri-continuous tri-crystalline structure. The colors represent
different lattice orientations. The characteristic distance $l$ between
the grains is indicated.}
\label{fig:Superlattices} 
\end{figure}

\begin{figure}[t]
\begin{centering}
\includegraphics[width=0.35\textwidth]{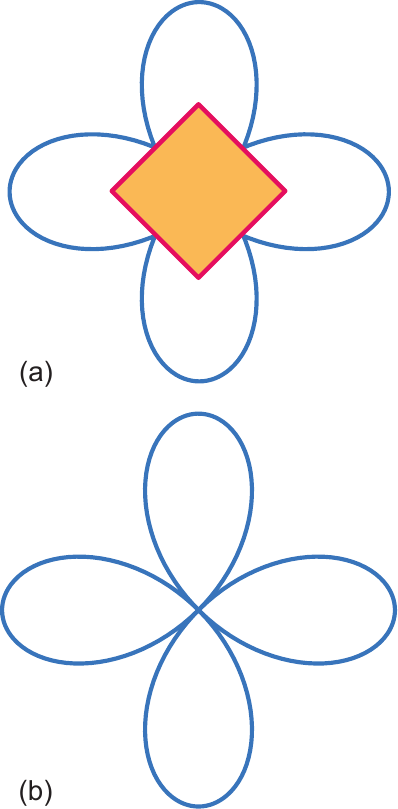} 
\par\end{centering}
\caption{2D schematic of possible $\gamma$-plots of a strongly anisotropic grain boundary. (a) Grain boundary with $\gamma>0$. The filled square shows the corresponding Wulff-Gibbs shape of an enclosed grain. (b) Grain boundary with $\gamma=0$.}
\label{fig:Gamma-plots} 
\end{figure}

The answer depends on many factors, such as the crystallographic anisotropy
of the boundary free energies, the temperature, and the free energy
cost of small deviations from the strict $\gamma=0$ condition. At
fixed temperature and pressure, up to $\xi=k-1$ grain boundaries can be
equilibrated. Some possible morphologies are shown in Fig.\ \ref{fig:Superlattices}.
If $\xi=1$ and the $\gamma=0$ condition can be satisfied by a symmetrical
tilt grain boundary, then the structure can be composed of lamellas
with twin-related crystallographic orientations (Fig.\ \ref{fig:Superlattices}(a)).
The spacing $l$ between the boundaries can be adjusted to match the
equilibrium area. This structure contains no extended defects other
than the grain boundaries and their intersections with the surface,
which can be neglected for a sufficiently large sample size. The lamellas
can also be organized into domains with orientations compatible with
the crystal symmetry (Fig.\ \ref{fig:Superlattices}(b)). The domain
boundaries in this polysynthetic structure are additional extended
defects, but their excess free energy can be neglected if they have
a large (e.g., mesoscopic) size. For $\xi>1$, the domains can be
composed of periodic arrangements of lamellas with different
crystallographic orientations ($\le\xi$) and widths adjusted to the required
specific area. In this case, the grain boundaries need not be symmetric.
Conventionally, lamella-based structures are vulnerable to capillary fluctuations
and morphological instabilities (such as the Plateau-Rayleigh instability)
unless the boundary free energy is highly
anisotropic with deep cusps at the inclinations represented in the
lamellas (Fig.\ \ref{fig:Gamma-plots}(a)). But these instabilities would be suppressed for a fully stabilized grain boundary, for which the cusps in the $\gamma$-plot reach the origin of the plot, at which $\gamma=0$, as shown schematically in Fig.\ \ref{fig:Gamma-plots}(b). 

An alternative morphology is a set of faceted grains embedded in a
large matrix grain. When $\xi=1$, the facets must have the same structure
up to symmetry operations (Fig.\ \ref{fig:Superlattices}(c)), while
for $\xi>1$ the facets can be different and the grains can have more
complex shapes (Fig.\ \ref{fig:Superlattices}(d)). The grains do
not have to be the same size. Like in the lamella case, strong crystallographic
anisotropy is required to avoid morphological instabilities at elevated
temperatures. The facet edges are additional extended defects with
excess free energy. They can create a capillary pressure acting on
the grains if $\gamma>0$. However, their contribution to the total
free energy can be negligible if the grains are sufficiently large.
Also, when the system is in the ground state, the grain boundaries at special misorientations/inclinations 
have zero tension (Fig.\ \ref{fig:Gamma-plots}(b)), and the force balance at the edges is of no concern.

At temperatures above the roughening transition, the excess free energy
of grain boundaries becomes nearly isotropic. In this case, the uniform
boundary model becomes a reasonable approximation, and all boundaries
can be treated as one 2D phase. The embedded grains need not be faceted.
If $\gamma>0$, the embedded grain shape must be close to spherical.
When $\gamma\rightarrow0$, the capillary forces vanish and the grains
have no particular shape or size. They may have a wide distribution
of sizes and ameaba-like shapes (Fig.\ \ref{fig:Superlattices}(e)).
Another possible morphology is a maze crystal, which can be considered
a particular case of the embedded-grain structure in which the embedded
grain and the matrix form interpenetrating pathways. More generally,
bi-continuous (Fig.\ \ref{fig:Superlattices}(f)), tri-continuous
(Fig.\ \ref{fig:Superlattices}(g)), and similar interpenetrating
morphologies are candidate ground-state structures when interfaces
are nearly isotropic. They contain no triple junctions and can easily
adjust the interface area per unit volume. 

Yet another possible scenario that arises at high temperatures is that $\gamma$ fluctuates around a zero value. The system will then reach
a dynamic equilibrium between grain growth (when $\gamma>0$) and
grain refinement (when $\gamma<0$) while maintaining a constant average
boundary area. Such dynamic grain structures were observed in recent
computer simulations \citep{Hussein:2025aa}.  In this case, the contribution of the configurational entropy to the free energy of the extended defects may no longer be negligible. Indeed, 1D defects could be analogous to polymer chains and 2D defects could be analogous to suspended graphene or lipid bilayers, all of which have configurational entropy contributions to free energy. After time-averaging over the fluctuations, the excess free energy may be zero.

The search for equilibrium microstructures that contain fully stabilized
grain boundaries is a design problem that becomes especially interesting
in the presence of interactions between the 1D/2D defects. So far, we have assumed that the defect excess free energies are simply additive, ignoring the interaction terms that depend on geometry.  Dislocations have a long-range stress field $\propto 1/r$, and even fully relaxed grain boundaries have a stress field that decays exponentially with distance from the boundary plane. In finite-size polycrystals, (weak) interactions are unavoidable.  For
microstructures composed of isolated defects, such as those shown
in Fig.\ \ref{fig:Superlattices}(c,d,e), the design problem reduces
to the optimization of a ``superlattice'' of 3D grains with a mesoscale
lattice parameter $l$. With enough chemical complexity, it seems
reasonable to expect that such a structure can reach a global minimum
at an optimal value of $l$. This ground-state superlattice structure will resist
grain coarsening or grain refinement, forming a "pseudo crystal" with internal interactions analogous to a ground-state atomic crystal. (Even though a ground-state atomic crystal has zero total stress, individual atomic pairs interact with attractive or repulsive interatomic forces, according to the Lennard-Jones interaction potential.)

As temperature or chemical composition change, this "pseudo crystal" structure can undergo reversible changes, assuming that
the solute diffusion is fast enough to respond to the changes. The
grain sizes and shapes can change, the lattice parameter $l$ can
increase (``thermal expansion'') or decrease, and the superlattice
itself can undergo sudden changes in the morphology and symmetry in
a manner similar to structural phase transformations of the atomic crystals. At a high temperature, the superlattice can ``melt'' by transitioning to a spatially disordered arrangement of the grains.

\section{Discussion and conclusions}

Previous thermodynamic treatments of extended defects as low-dimensional
phases considered only phase equilibria and phase transformations
\emph{within} a defect with a fixed area or length \citep{Frolov:2015ab}.
Thus, only constrained thermodynamic equilibria of the entire system
were considered. The defects were in thermal, mechanical, and chemical
equilibrium with their environment, but not in full equilibrium. The
system containing extended defects was always in a metastable state.
In this paper, we continued to treat extended defects as low-dimensional
phases. However, we applied this treatment to address new questions
not asked before, such as: Can a crystalline material with multiple
extended defects reach a true thermodynamic equilibrium? If it can,
what thermodynamic properties will the defect phases have? What kind
of microstructure will the material have when it reaches the thermodynamic
ground state?

To let the system reach the full thermodynamic equilibrium, we allowed
the extended defects to vary their areas and lengths, which are the
degrees of freedom that were previously frozen. In other words, the
amounts of phases of \emph{all} dimensionalities in a closed system
were allowed to vary to reach full equilibrium. Two different scenarios
were found. The defect phases can shrink in size and eventually disappear,
leaving only bulk phases in the ground state. This happens if the
defect formation free energies remain positive during the equilibration
process. Grain growth in polycrystalline alloys is an example of this
process. However, thermodynamics also permits a scenario in which
some special extended defects remain. We have shown that the formation
free energies of such defects in the ground state must be zero. In
fact, such special "defects" are no longer defects but low-dimensional phases
that can coexist thermodynamically with the conventional 3D phases. All driving forces for
microstructure coarsening vanish, and the material becomes structurally
stable. An example is offered by the hypothetical fully stabilized
nanocrystalline alloys discussed in the literature \citep{Weissmuller1994,Weissmuller:1993aa,Kirchheim:2002aa,Liu:2004aa,Krill:2005aa,Schvindlerman2006,Kirchheim2007a,Kirchheim2007b,Detor2007,Trelewicz2009,Chookajorn2012,Murdoch:2013aa,Chookajorn2014,Kalidindi:2017aa,Kalidindi:2017bb,Kalidindi:2017cc,Perrin-2021,Hussein:2024aa,Hussein:2025aa}.
However, many other types of stabilized materials, e.g., "pseudo crystals", can be imagined
with unique properties in the realms of ceramics and electronic materials.  These structures with a finite characteristic length scale are forever immune to coarsening if the temperature and chemical composition are fixed. However, the said length scale
can also vary when the intensive thermodynamic variables are changed, akin to thermal expansion or chemical expansion of atomic crystals. 

To the best of our knowledge, full thermodynamic stabilization of polycrystals or any other defected structures has not been demonstrated experimentally so far. One of the challenges in achieving the full stabilization in polycrystalline structures is that grain boundary segregation thermodynamically competes with bulk phase transformations. Rather than forming strong GB segregation atmospheres that could stabilize the polycrystalline state, the solute atoms may prefer to precipitate as a new bulk phase. However, it is conceivable that in some cases the bulk phase nucleation can be delayed by slow kinetics, and the grain boundaries in the initial bulk phase can still reach a free-energy minimum at a finite grain size. Although this state would be thermodynamically metastable, the phase rule formulated above would still apply to the stabilized grain boundaries and other extended defects within the initial phase. Of course, on a much longer timescale, the solute atoms would gradually diffuse away from the grain boundaries into the new phase, eventually causing grain coarsening and destabilization of other extended defects. 

In the beginning of the article, we mentioned intrinsic point defects whose free energy includes a configurational part due to their mobility and which are part of a bulk phase. However, many microstructures contain immobile point defects, such as dislocation nodes and quadruple points in polycrystalline materials. Such 0D defects (``defects in defects'') should be considered as 0D phases, and their degrees of freedom should be included in the phase rules. Considering that the fraction of atoms that reside in such phases is relatively small, we focused on 1D and 2D defects. Extension of our analysis to include 0D phases is straightforward. 

An important outcome of this work is the realization that the stabilized
extended defects must follow Gibbs' phase rule generalized to phases
of any dimensionality. This rule limits the number of defect types
that the material can contain in the ground state. This number depends
on the number of chemical components in the system and is in practice
relatively small. The phase rule can serve as a guide for the design
of thermodynamically stabilized materials. It also opens up an exciting
new direction of searching for microstructures that contain a prescribed
number of extended defects and are capable of adjusting the defect
areas/length to satisfy the equilibrium conditions. We discussed several
possible microstructures for a single-phase solid solution with grain
boundaries. However, this simple case barely scratches the surface
of the problem. Deeper and more systematic investigations are needed
in the future.

Many other aspects of full stabilization call for future research.
For example, solute diffusion is a critical factor that was not discussed
here. During the equilibration process, the solute diffusion must
be fast enough to sustain the segregation atmospheres at the defects
and to maintain the low and eventually zero values of their formation free
energies. This is especially important for the dynamically equilibrated
structures in which the defects constantly move and their free energies
fluctuate between positive and negative values. Recent simulations
suggest that slow solute diffusion can trap the structure in metastable
states \citep{Hussein:2024aa}.

Another unexplored question is how the ground-state structures with
defects would appear on phase diagrams. In the simple case of a binary
solid solution (phase $\alpha$) with a single grain boundary ($(1@1@0)_{2}$
system), the ground state structure can exist in a temperature-composition
domain bounded by coexistence lines with other phases (Fig.\ \ref{fig:diagrams-B}(c)).
At a fixed temperature, the ground states span the composition interval
between the end points $a$ and $b$ on the tie line. Near point $a$,
the grain boundary area is infinitely small and phase $\alpha$ is
virtually a single crystal. Near point $b$, most atoms belong to
the grain boundary while the grains are infinitely small. In reality,
phase $\alpha$ is likely to transform to another bulk phase before
point $b$ can be reached. Alternatively, the grain boundary can premelt
when approaching point $b$ and then fully melt at this point. In
this scenario, the domain of the $(1@1@0)_{2}$ system on the phase
diagram is bounded by a solid-liquid coexistence field on the high-concentration
side. In recent computer simulations \citep{Hussein:2025aa}, a stable
polycrystalline state in a binary system was indeed observed under
a solidus line on the phase diagram. Future research can show how
general this trend is and what fully stabilized states can look like
on the structural phase diagrams of more complex systems.

\vspace{0.15in}

\textbf{Acknowledgement:} J.~L.~acknowledges support of the National Science
Foundation, Division of Materials Research, under Award DMR-1923976. Y.~M.~was supported by the National Science Foundation, Division of Materials Research, under Award DMR-1708314. 


\newpage \clearpage

 {\title{\bf \centering{SUPPLEMENTARY INFORMATION}}} \\

\bigskip
{\large    Line and planar defects with zero formation free energy: Applications of the phase rule toward ripening-immune microstructures}\\

\bigskip
\author{Ju Li$^{1,2}$ and Yuri Mishin$^{3}$}
\maketitle

\lyxaddress{$^{1}$ Department of Nuclear Science and Engineering, Massachusetts
    Institute of Technology, Cambridge, MA 02139, USA}

\lyxaddress{$^{2}$ Department of Materials Science and Engineering, Massachusetts
    Institute of Technology, Cambridge, MA 02139, USA}

\lyxaddress{$^{3}$ Department of Physics and Astronomy, MSN 3F3, George Mason
    University, Fairfax, Virginia 22030, USA}

\bigskip

\noindent In this Supplementary Information file, we show that the condition
of zero formation free energies of all extended defects is equivalent
to the condition of thermodynamic equilibrium among all 1D, 2D, and
3D phases present in the system. To simplify the derivation, we will
assume that the phases have the same temperature at all times. In
other words, we take for granted that the phases always remain in
thermal equilibrium with each other.

A spatially uniform 3D phase composed of $k$ chemical components
is fully described by a single equation expressing its Helmholtz free
energy as a function of temperature $T$, volume $V$, and the amounts
$N_{i}$ of the chemical components (measured in moles): 
\begin{equation}
    F=F(T,N_{1},...,N_{k},V).\label{eq:10}
\end{equation}
Gibbs \citep{Gibbs} referred to such equations as ``fundamental''
because they fully define all thermodynamic properties of the system.\footnote{Gibbs used a more general fundamental equation in the form $S=S(E,N_{1},...,N_{k},V)$,
    $S$ being the system entropy and $E$ the system energy. Since we
    assume thermal equilibrium across the system, we can conveniently
    reformulate the fundamental equation in terms of the free energy,
    as in Eq.(\ref{eq:10}).} Such properties can be calculated by straightforward calculus without
any additional information about the system.

The fundamental equation (\ref{eq:10}) is a homogeneous function
of first degree with respect to the extensive variables $N_{1},...,N_{k}$
and $V$. Applying the Euler theorem of homogeneous functions, we
obtain 
\begin{equation}
    F=\sum_{i=1}^{k}\mu_{i}N_{i}-pV,\label{eq:11}
\end{equation}
where $\mu_{i}=\partial F/\partial N_{i}$ are the chemical potentials
of the components and $p=-\partial F/\partial V$ is pressure. On
the other hand, taking the differential of Eq.(\ref{eq:10}) we have
\begin{equation}
    dF=-SdT+\sum_{i=1}^{k}\mu_{i}dN_{i}-pdV,\label{eq:12}
\end{equation}
where $S=-\partial F/\partial T$ is the system entropy.

Now consider an interface separating two phases or two grains within
the same phase. It can be shown \citep{Frolov:2015ab} that all interface
properties are fully defined by the fundamental equation 
\begin{equation}
    \tilde{F}=\tilde{F}(T,\tilde{N}_{1},...,\tilde{N}_{k},A),\label{eq:14}
\end{equation}
where $A$ is the interface area, $\tilde{F}$ is the excess free
energy of the interface, and $\tilde{N}_{i}$ are the excess amounts
(in moles) of the chemical components. The excesses are defined using
the dividing surface construction \citep{Gibbs}, in which the excess
is taken relative to intensive properties of the homogeneous bulk
phases extrapolated to the dividing surface.\footnote{Using the dividing surfaces simplifies thermodynamic derivations without
    loss of generality. All final results can be readily reformulated
    in terms of Cahn's generalized excess formalism \citep{Cahn79a} by
    mathematical rearrangements \citep{Frolov:2015ab}.} Note that by deriving the interface thermodynamics starting from
a fundamental equation, such as Eq.(\ref{eq:14}), we treat interfaces
as 2D phase. This interpretation is consistent with Gibbs \citep{Gibbs},
who treated all systems described by a fundamental equation the same
way as he treated bulk phases.

Eq.(\ref{eq:14}) has the same functional form as Eq.(\ref{eq:10})
for bulk phases, except that the spatial dimensions of the interface
are defined by its area $A$ instead of the volume $V$. The excess
free energy in Eq.(\ref{eq:14}) is again a homogeneous function of
first degree. Therefore, we can immediately rewrite the above equations
in the interface variables: 
\begin{equation}
    \tilde{F}=\gamma A+\sum_{i=1}^{k}\tilde{\mu_{i}}\tilde{N}_{i},\label{eq:23}
\end{equation}
\begin{equation}
    d\tilde{F}=-\tilde{S}dT+\gamma dA+\sum_{i=1}^{k}\tilde{\mu_{i}}d\tilde{N}_{i},\label{eq:24}
\end{equation}
where $\tilde{S}=-\partial\tilde{F}/dT$ is the excess entropy. We
denoted $\tilde{\mu_{i}}\equiv\partial\tilde{F}/\partial\tilde{N}_{i}$
and defined 
\begin{equation}
    \gamma=\dfrac{\partial\tilde{F}}{\partial A}\label{eq:gamma_1}
\end{equation}
as the interface free energy (interface tension).

Similarly, we can derive all thermodynamic properties of a 1D defect
of length $L$ from the fundamental equation 
\begin{equation}
    \hat{F}=\hat{F}(T,\hat{N}_{1},...,\hat{N}_{k},L),\label{eq:15}
\end{equation}
which again implies that we treat the 1D defect as a phase. The excess
free energy in Eq.(\ref{eq:15}) is a homogeneous function of first
degree similar to Eq.(\ref{eq:10}) except for the replacement of
$V$ by $L$. As above, we immediately write down the equations 
\begin{equation}
    \hat{F}=\tau L+\sum_{i=1}^{k}\hat{\mu}_{i}\hat{N}_{i},\label{eq:16}
\end{equation}
\begin{equation}
    d\hat{F}=-\hat{S}dT+\tau dL+\sum_{i=1}^{k}\hat{\mu}_{i}d\hat{N}_{i},\label{eq:17}
\end{equation}
where $\hat{S}=-\partial\hat{F}/dT$ is the excess entropy of the
1D defect. We denoted $\hat{\mu}_{i}\equiv\partial\hat{F}/\partial\hat{N}_{i}$
and defined 
\begin{equation}
    \tau=\dfrac{\partial\hat{F}}{\partial L}\label{eq:gamma_1-1}
\end{equation}
as the 1D defect free energy (line tension).

Next, we consider a compound system composed of $\varphi_{1}$ 1D
defects, $\varphi_{2}$ 2D defects, and $\varphi_{3}$ 3D phases.
Adding the above equations, we obtain the total free energy of the
system, 
\begin{eqnarray}
    F_{\mathrm{tot}} & = & \sum_{l=1}^{\varphi_{1}}\left(\tau_{l}L_{l}+\sum_{i=1}^{k}\hat{\mu}_{i}^{(l)}\hat{N}_{i}^{(l)}\right)\nonumber \\
    & + & \sum_{m=1}^{\varphi_{2}}\left(\gamma_{m}A_{m}+\sum_{i=1}^{k}\tilde{\mu}_{i}^{(m)}\tilde{N}_{i}^{(m)}\right)\nonumber \\
    & + & \sum_{n=1}^{\varphi_{3}}\left(-p_{n}V_{n}+\sum_{i=1}^{k}\mu_{i}^{(n)}N_{i}^{(n)}\right),\label{eq:21}
\end{eqnarray}
and its differential 
\begin{eqnarray}
    dF_{\mathrm{tot}} & = & \sum_{l=1}^{\varphi_{1}}\left(-\hat{S}_{l}dT+\tau_{l}dL_{l}+\sum_{i=1}^{k}\hat{\mu}_{i}^{(l)}d\hat{N}_{i}^{(l)}\right)\nonumber \\
    & + & \sum_{m=1}^{\varphi_{2}}\left(-\tilde{S}_{m}dT+\gamma_{m}dA_{m}+\sum_{i=1}^{k}\tilde{\mu}_{i}^{(m)}d\tilde{N}_{i}^{(m)}\right)\nonumber \\
    & + & \sum_{n=1}^{\varphi_{3}}\left(-S_{n}dT-p_{n}dV_{n}+\sum_{i=1}^{k}\mu_{i}^{(n)}dN_{i}^{(n)}\right).\label{eq:22}
\end{eqnarray}

\begin{flushleft}
    Suppose the system is closed (no matter exchange with the environment)
    and its volume and temperature are fixed. These constraints can be
    formulated by 
    \begin{equation}
        \sum_{l=1}^{\varphi_{1}}d\hat{N}_{i}^{(l)}+\sum_{m=1}^{\varphi_{2}}d\tilde{N}_{i}^{(m)}+\ \sum_{n=1}^{\varphi_{3}}d{N}_{i}^{(n)}=0,\;i=1,..,k,\label{eq:19}
    \end{equation}
    \begin{equation}
        \sum_{n=1}^{\varphi_{3}}dV_{n}=0.\label{eq:20}
    \end{equation}
    A system subject to these constraints spontaneously decreases $F_{\mathrm{tot}}$
    until it reaches a global minimum in the equilibrium state. The necessary
    condition of the minimum is 
    \begin{eqnarray}
        (dF_{\mathrm{tot}})_{T} & + & \sum_{i=1}^{k}\lambda_{i}\left(\sum_{l=1}^{\varphi_{1}}d\hat{N}_{i}^{(l)}+\sum_{m=1}^{\varphi_{2}}d\tilde{N}_{i}^{(m)}+\ \sum_{n=1}^{\varphi_{3}}d{N}_{i}^{(n)}\right)\nonumber \\
        & + & \omega\sum_{n=1}^{\varphi_{3}}dV_{n}=0,\label{eq:26}
    \end{eqnarray}
    where $\lambda_{i}$ and $\omega$ are Lagrange multipliers for the
    constraints (\ref{eq:19}) and (\ref{eq:20}), respectively. Inserting
    Eq.(\ref{eq:22}) for $dF_{\mathrm{tot}}$, the minimum condition
    becomes 
    \begin{alignat}{1}
        & -\sum_{n=1}^{\varphi_{3}}(p_{n}-\omega)dV_{n}+\sum_{l=1}^{\varphi_{1}}\tau_{l}dL_{l}+\sum_{m=1}^{\varphi_{2}}\gamma_{m}dA_{m}\nonumber \\
        & +\sum_{l=1}^{\varphi_{1}}\left(\sum_{i=1}^{k}\hat{\mu}_{i}^{(l)}+\lambda_{i}\right)d\hat{N}_{i}^{(l)}+\sum_{m=1}^{\varphi_{2}}\left(\sum_{i=1}^{k}\tilde{\mu}_{i}^{(m)}+\lambda_{i}\right)d\tilde{N}_{i}^{(m)}\nonumber \\
        & +\sum_{m=1}^{\varphi_{3}}\left(\mu_{i}^{(n)}+\lambda_{i}\right)dN_{i}^{(n)}=0,\label{eq:27}
    \end{alignat}
    where all differentials are considered independent variations. When
    the minimum is reached, all differential coefficients vanish. 
    \par\end{flushleft}

\begin{flushleft}
    The mechanical equilibrium condition is obtained by setting the differential
    coefficients in front of the $dV_{n}$ variations to zero, which gives
    \begin{equation}
        p_{1}=p_{2}=...=p_{\varphi_{3}}.\label{eq:28}
    \end{equation}
    All bulk phases are at the same pressure. 
    \par\end{flushleft}

\begin{flushleft}
    The chemical equilibrium condition is obtained from the terms with
    the differentials of the amounts of chemical components, which gives
    \begin{equation}
        \mu_{i}^{(1)}=\mu_{i}^{(2)}=...=\mu_{i}^{(\varphi_{3})}\equiv\mu_{i},\;i=1,..,k,\label{eq:29}
    \end{equation}
    \begin{equation}
        \tilde{\mu}_{i}^{(1)}=\tilde{\mu}_{i}^{(2)}=...=\tilde{\mu}_{i}^{(\varphi_{2})}=\mu_{i},\;i=1,..,k,\label{eq:30}
    \end{equation}
    \begin{equation}
        \hat{\mu}_{i}^{(1)}=\hat{\mu}_{i}^{(2)}=...=\hat{\mu}_{i}^{(\varphi_{1})}=\mu_{i},\;i=1,..,k.\label{eq:31}
    \end{equation}
    In other words, the chemical potential of every component is the same
    in all 1D, 2D, and 3D phases. All segregation atmospheres are in equilibrium
    with the surrounding bulk regions with respect to exchanges of chemical
    components. 
    \par\end{flushleft}

\begin{flushleft}
    Once the mechanical and chemical equilibrium conditions are satisfied,
    the free energy minimum condition (\ref{eq:27}) reduces to 
    \begin{equation}
        \sum_{l=1}^{\varphi_{1}}\tau_{l}dL_{l}+\sum_{m=1}^{\varphi_{2}}\gamma_{m}dA_{m}=0.\label{eq:32}
    \end{equation}
    At this point, the 3D phases are in equilibrium with each other, but
    the defect phases still remain in constrained equilibria. The full
    equilibration of the system requires that the total free energy be
    a minimum with respect to arbitrary variations of the lengths and
    areas of all extended defects. Such variations are represented by
    the differentials $dL_{l}$ and $dA_{m}$. Thus the full equilibration
    requires that the defect free energies be zero: 
    \begin{equation}
        \tau_{1}=\tau_{2}=...=\tau_{\varphi_{1}}=0,\label{eq:33}
    \end{equation}
    \begin{equation}
        \gamma_{1}=\gamma_{2}=...=\gamma_{\varphi_{2}}=0.\label{eq:34}
    \end{equation}
    These are the conditions we sought to prove. 
    \par\end{flushleft}

Returning to the case where the extended defects are in constrained
but not full equilibrium, Eqs.(\ref{eq:16}) and (\ref{eq:23}) give
the following equations for the excess free energies of the extended
defects: 
\begin{equation}
    \hat{F}_{l}=\tau_{l}L_{l}+\sum_{i=1}^{k}\mu_{i}\hat{N}_{i}^{(l)},\;l=1,..,\varphi_{1},\label{eq:35}
\end{equation}
\begin{equation}
    \tilde{F}_{m}=\gamma_{m}A_{m}+\sum_{i=1}^{k}\mu_{i}\tilde{N}_{i}^{(m)},\;l=1,..,\varphi_{2}.\label{eq:36}
\end{equation}
Accordingly, the total free energy of a solid with metastable defects
is 
\begin{equation}
    F_{\mathrm{tot}}=\sum_{l=1}^{\varphi_{1}}\tau_{l}L_{l}+\sum_{m=1}^{\varphi_{2}}\gamma_{m}A_{m}+\sum_{i=1}^{k}\mu_{i}N_{i}-pV,\label{eq:37}
\end{equation}
where $N_{i}$ are the total amounts of the components in the system.
In the thermodynamic ground state, 
\begin{equation}
    F_{\mathrm{tot}}=\sum_{i=1}^{k}\mu_{i}N_{i}-pV.\label{eq:38}
\end{equation}
The ground-state free energy follows the same equation as a single-phase
solid.

It is now easy to derive the phase rule for an equilibrium multiphase
system with extended defects. According to the fundamental equations
(\ref{eq:10}), (\ref{eq:14}) and (\ref{eq:15}), each phase is characterized
by $k+1$ intensive parameters. The total number of intensive parameters
in all phases is $(\varphi_{1}+\varphi_{2}+\varphi_{3})(k+1)$. The
three equilibrium conditions (\ref{eq:29}), (\ref{eq:30}) and (\ref{eq:31})
plus the equality of temperatures in all phases impose 
\begin{equation}
    (k+1)\left[\varphi_{1}+\varphi_{2}+\varphi_{3}-1\right]
\end{equation}
constraints. Eqs.(\ref{eq:28}), (\ref{eq:33}) and (\ref{eq:34})
impose additional 
\begin{equation}
    \varphi_{1}+\varphi_{2}+(\varphi_{3}-1)
\end{equation}
constraints, bringing the total number of constraints to 
\begin{equation}
    (k+2)(\varphi_{1}+\varphi_{2}+\varphi_{3})-k-2.
\end{equation}
The number of degrees of freedom of the system, $\pi$, is the number
of intensive parameters minus the number of constraints, which gives
\begin{equation}
    \pi=k+2-(\varphi_{1}+\varphi_{2}+\varphi_{3}).\label{eq:39}
\end{equation}

\end{document}